\documentclass[fleqn,10pt]{wlscirep}
\usepackage[utf8]{inputenc}
\usepackage[T1]{fontenc}
\usepackage{braket}
\title{Near-optimal discrimination of displaced squeezed binary signals using displacement, inverse-squeezing, and photon-number-resolving detection}

\author[1,2]{Enhao Bai}
\author[1]{Jian Peng}
\author[1]{Tianyi Wu}
\author[2]{Kai Wen}
\author[2]{Fengkai Sun}
\author[3]{Chun Zhou}
\author[4]{Yaping Li}
\author[2,*]{Zhenrong Zhang}
\author[1,+]{Chen Dong}

\affil[1]{Information Support Force Engineering University, Wuhan 430035, China}
\affil[2]{Guangxi Key Laboratory of Multimedia Communications and Network Technology, Guangxi University, Nanning 530006, China}
\affil[3]{Henan Key Laboratory of Quantum Information and Cryptography, SSF IEU, Zhengzhou 450001, China}
\affil[4]{Wuhan Maritime Communication Research Institute, Wuhan 430035, China}

\affil[*]{Corresponding author, Email: zzr76@gxu.edu.cn}
\affil[+]{Corresponding author, Email: dongchengfkd@163.com}
\begin{abstract}
We propose an inverse-squeezing Kennedy receiver for discriminating binary phase-shift-keyed displaced squeezed vacuum states. The receiver combines a Kennedy-type nulling displacement, an orthogonally oriented inverse-squeezing operation and photon-number-resolving detection with a maximum-a-\emph{posteriori} threshold rule.
Its key mechanism is that the inverse-squeezing stage converts transmitter-side squeezing into enhanced photon-number contrast, or equivalently an effective coherent-state energy gain, that can be directly exploited at the measurement stage.
Under ideal equal-prior conditions, the receiver surpasses the standard quantum limit for squeezed-state binary phase-shift keying at approximately $N\approx 0.3$, outperforms the Helstrom bound of coherent-state binary phase-shift keying at approximately $N\approx 0.4$, and reaches the 1\% error level near $N\approx 0.6$.
We further analyze its performance under realistic imperfections, including finite detector efficiency, dark counts, channel phase diffusion, receiver thermal noise and transmission loss. The results show that adaptive thresholding preserves robust performance against detector and noise imperfections over practical parameter ranges, whereas transmission loss progressively suppresses the squeezing-enabled advantage. These findings indicate that, for the fixed source parametrization adopted in this work, the proposed receiver is most advantageous in the low-loss regime, especially at low source energies.
\end{abstract}
\begin{document}

\flushbottom
\maketitle
\thispagestyle{empty}

\section*{Introduction}
Quantum state discrimination is a central problem in quantum information and optical communications, because the minimum achievable detection error directly determines the reliability of information transfer. Among optical encodings, coherent states have long played a dominant role owing to their experimental accessibility and their robustness under bosonic loss \cite{coherent_state_01, coherent_state_02, coherent_state_03, review_2012}. This has led to the development of a wide range of receiver architectures for coherent-state discrimination, including the Kennedy receiver\cite{Kennedy_Receiver}, optimized-displacement receivers\cite{Optimal_Displacement}, multistage and feedforward receivers\cite{Partition_Adaptive_Nulling_Receiver, Shuro_Receiver, chen2015, Sequential_Waveform_Nulling_Receiver, L_M_CPN}, hybrid detection strategies \cite{Hybrid_Receiver, Homodyne_Like, Olivares_01, zuo_hyb}. Nevertheless, coherent states remain non-orthogonal, and hence their discrimination error probability is fundamentally bounded by the Helstrom bound (HB) \cite{Helstrom,SRM01}.

In parallel with these classical developments, recent years have witnessed substantial progress in coherent-state discrimination beyond the conventional binary signal. These advances include photon-counting and structured optical receivers approaching the quantum limit \cite{R1_03, R1_07}, receiver designs for multi-symbol and multi-mode constellations \cite{bai_qpsk_hybrid, sun_bpsk_multichannel, squeeze_2013, R1_02, R1_06}, and learning-assisted strategies for receiver optimization, calibration, and near-optimal joint detection \cite{R1_01, R1_04, R1_05}. Such studies have significantly broadened the scope of quantum receiver theory, both in terms of signal alphabets and measurement strategies. However, most of these developments still focus on coherent-state alphabets, i.e., on improving the receiver side while keeping the transmitted states classical in the phase-space sense.

A different route is to exploit nonclassical transmitter states whose intrinsic overlap is smaller than that of coherent states at the same energy \cite{squeeze_2001, Olivares2018, DSS_2025_arxiv}. In this context, Ref.~\cite{Olivares2018} showed that squeezed-vacuum-based phase-shift-keyed communication can outperform coherent-state binary phase-shift keying (BPSK) benchmarks. In particular, by displacing a squeezed vacuum state with opposite phases, one obtains the binary alphabet of phase-opposite displaced squeezed vacuum states,
\begin{equation}
  \label{eq:BPSK_DSS}
  \left\{\begin{aligned}
    \text{symbol}\ '0':\ &\left|\psi_0\right\rangle = \hat{D}(-\alpha) \hat{S}(r) \left|0\right\rangle = \left|-\alpha,r\right\rangle \\
    \text{symbol}\ '1':\ &\left|\psi_1\right\rangle = \hat{D}(+\alpha) \hat{S}(r) \left|0\right\rangle = \left|+\alpha,r\right\rangle
  \end{aligned}\right.
\end{equation}
which we refer to as squeezed-state BPSK (S-BPSK). For a fixed mean signal energy, the Helstrom bound of S-BPSK can be significantly lower than that of coherent-state BPSK (C-BPSK), and even the homodyne limit of S-BPSK may fall below the Helstrom bound of C-BPSK in the low-energy regime \cite{Olivares2018}. These results indicate that transmitter-side squeezing can serve as a genuine communication resource for error reduction.

Motivated by this observation, the key open question is how to design a practical receiver that can efficiently convert the squeezing resource of the transmitted S-BPSK signals into an experimentally accessible discrimination advantage. In this work, we address this problem by proposing the inverse-squeezing Kennedy receiver (IS-Kennedy), which combines a displacement operation, an orthogonally oriented inverse-squeezing (IS) operation, and photon-number-resolving (PNR) detection with a maximum-a-\emph{posteriori} (MAP) threshold rule. The basic idea is to map the incoming displaced squeezed-state alphabet into a more distinguishable effective coherent-state on--off-keying pair, thereby increasing the phase-space separation and reducing the overlap of the corresponding photon-number statistics. Under ideal conditions, we show that the error probability of the proposed receiver always remains within a factor of two of the S-BPSK Helstrom bound for equal priors, while outperforming coherent-state benchmarks in the low-energy regime. We further analyze its robustness under realistic imperfections, including finite detector efficiency, dark counts, phase diffusion, receiver thermal noise, and transmission loss.

\section*{Results}
\subsection*{Channel model and performance benchmarks}
In this section we summarize the signal model and the fundamental performance limits for BPSK communication based on displaced squeezed vacuum states, and we fix the notation used in the rest of the paper.

For binary discrimination between quantum states $\left|\psi_0\right\rangle$ and $\left|\psi_1\right\rangle$ with prior probabilities $\pi_0 = \pi_1 = \frac{1}{2}$, the minimum error probability (Helstrom bound) is given by \cite{Helstrom, Olivares2018, review_2021}:
\begin{equation}
  \label{eq:Helstrom_general}   
  \begin{aligned}
    P_{\text{HB}}^{\text{S-BPSK}}\left(\alpha, r\right) 
    &= \frac{1}{2} \left(1 - \sqrt{1 - \left|\left\langle \psi_0|\psi_1 \right\rangle\right|^2}\right) \\
    &= \frac{1}{2} \left(1 - \sqrt{1 - \exp\left(-4\alpha^2\text{e}^{2r}\right)}\right)
  \end{aligned}
\end{equation}

As a practically relevant benchmark we also consider homodyne detection along the X quadrature, which attains the standard quantum limit (SQL) for BPSK under ideal conditions. Given the input is $\rho_i = \left|\psi_i\right\rangle\left\langle\psi_i\right|$, the conditional probability density of the homodyne outcome $x$ ,
\begin{equation}
  \label{eq:homodyne_probability}
  \begin{aligned}
    &P\left( x|\rho_i \right) = \int_{-\infty}^{+\infty} \text{d}p \cdot W\left(x,p|\rho_i\right) \\
    &= \frac{\mathrm{e}^{r}}{\sqrt{\pi}}\, \exp\left\{ -\,\mathrm{e}^{2r}\left[ x-(-1)^{i+1} \sqrt{2}\alpha\right]^2 \right\},\ i\in{0,1},
  \end{aligned}
\end{equation}
which $W\left(x,p|\rho_i\right)$ is the Wigner function of $\rho_i$ (see \emph{Supplementary Information I}). And the resulting error probability is
\begin{equation}
  \label{eq:SQL}
  P_{\text{SQL}}^{\text{S-BPSK}} = \int_{0}^{+\infty} P\left( x|\rho_0 \right) \mathrm{d}x
  = \frac{1}{2} \text{erfc}\left(\sqrt{2}\alpha \text{e}^r\right)
\end{equation}

To parameterize the contribution of squeezing, we introduce the squeezing fraction
\begin{equation}
  \beta \triangleq \frac{\sinh^2(r)}{N}
\end{equation}
which quantifies the share of the total energy ($N=\left|\alpha\right|^2 + \sinh^2(r)$) stored in squeezing \cite{Olivares2018}. Therefore, the parameter $\alpha$ and $r$ can be expressed as
\begin{equation}
  \left\{
  \begin{aligned}
    &\alpha = \sqrt{N\left(1 - \beta\right)}, \quad r = \sinh^{-1}\sqrt{N\beta} \\
    &\alpha\text{e}^r = \sqrt{N\left(1 - \beta\right)}\left(\sqrt{1 + N\beta} + \sqrt{N\beta}\right)
  \end{aligned}\right.
\end{equation}
Accordingly, both the Helstrom bound and the SQL for S-BPSK become functions of $\left(N, \beta\right)$, denoted $P_{\text{SQL}}^{\text{S-BPSK}}\left(N,\beta\right)$, and $P_{\text{HB}}^{\text{S-BPSK}}\left(N,\beta\right)$.
The squeezing fraction can be optimized at fixed total energy. By minimizing $P_{\text{HB}}^{\text{S-BPSK}}\left(N,\beta\right)$ over $\beta$ one obtains the optimal value \cite{Olivares2018}:
\begin{equation}
  \label{eq:optimal_beta}
  \beta_{\text{opt}}(N) = \frac{N}{2N + 1}, \quad \alpha \text{e}^{r} = \sqrt{N(N+1)}
\end{equation}
Both the Helstrom bound and the SQL for S-BPSK are rewritten as:
\begin{align}
  &P_{\text{SQL}}^{\text{S-BPSK}}\left(N,\beta_{\text{opt}}\right) = \frac{1}{2}\text{erfc}\left(\sqrt{2N\left(N+1\right)}\right) \label{eq:DSS_SQL_opt}\\
  &P_{\text{HB}}^{\text{S-BPSK}}\left(N,\beta_{\text{opt}}\right) = \frac{1}{2} \left(1 - \sqrt{1 - \exp\left\{-4N(N+1)\right\}}\right) \label{eq:DSS_HB_opt}
\end{align}

\begin{figure}[htbp]
  \centering
  \includegraphics[scale=1.0]{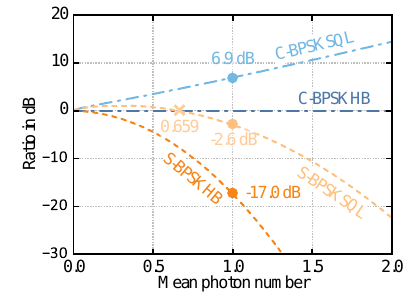}
  \caption{Comparison of the Helstrom bound (HB) and the standard quantum limit (SQL) for error probabilities in binary phase-shift keying (BPSK) with displaced squeezed states (S-BPSK) and coherent states (C-BPSK). Error probability is plotted against the signal mean photon number $N$. The C-BPSK Helstrom bound $P_\text{HB}^\text{C-BPSK} = \frac{1}{2} \left[ 1- \sqrt{1-\exp\left (-4\cdot N \right )} \right]$ is used as the benchmark, and the corresponding standard quantum limit is $P_\text{SQL}^\text{C-BPSK} = \frac{1}{2} \left[ 1-\mathrm{erf}\left( \sqrt{2\cdot N} \right) \right]$.} 
  \label{fig:Gain_HB_SQL}
\end{figure}

When the S-BPSK are prepared with this choice, the S-BPSK Helstrom bound is significantly lower than that of C-BPSK at the same energy. In particular, for $N\approx1$ the S-BPSK Helstrom bound already reaches the order of $10^{-4}$, whereas C-BPSK requires approximately twice the energy to achieve a comparable error probability. Moreover, in the low-photon number regime the S-BPSK SQL can fall below the Helstrom bound of C-BPSK, with the crossover occurring around $N\approx0.659$ (see Fig.~\ref{fig:Gain_HB_SQL}). At $N = 1.0$, the Helstrom bound for S-BPSK surpasses the SQL for C-BPSK by a remarkable 17.39dB, and the SQL for S-BPSK also is 2.94dB lower. In the subsequent analyses, all S-BPSK employ the optimal squeezing fraction $\beta_{\text{opt}}$.

\subsection*{Inverse-squeezing Kennedy receiver under ideal conditions}

\begin{figure}[htbp]
  \centering
  \includegraphics[scale=1.0]{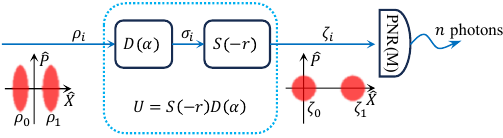}
  \caption{Schematic of the proposed inverse-squeezing Kennedy receiver. The input state $\rho_i$ is first processed by a Kennedy-type nulling displacement $D(\alpha)$, which maps the BPSK displaced squeezed states to a squeezed on-off keying pair. A subsequent inverse-squeezing operation $S(-r)$ converts the two states from the displaced squeezed alphabet into an amplified coherent-state alphabet, ideally $\{\ket{0},\;\ket{2\alpha e^{r}}\}$ in the matched case. The resulting state $\zeta_i$ is measured by a photon-number-resolving (PNR) detector yielding outcome $n$, and the final decision is made according to the maximum a \emph{posteriori} criterion.}
  \label{fig:structure_of_ISK}
\end{figure}

In this section, we propose a novel quantum-enhanced receiver for S-BPSK discrimination, referred to as the IS-Kennedy. Its schematic is shown in Fig.~\ref{fig:structure_of_ISK}. 
Compared to the conventional Kennedy receiver, the IS-Kennedy incorporates two key modifications: \emph{(i)} the insertion of an IS module after the displacement operation, which applies the operator $S(-r)$ to the signal; and \emph{(ii)} the upgrade of the terminal single-photon detector (SPD) to a photon-number-resolving (PNR) detector, enabling MAP decision-making under generalized statistical models. The squeezing amplitude of the IS module is matched to that of the transmitter, while the squeezing axis is rotated by $\pi/2$ (orthogonal) relative to the transmitter. This configuration cancels the initial squeezing and produces an equivalent amplification of the signal's mean photon number. Effectively, the IS-Kennedy converts the transmitter's squeezing resources into an equivalent amplitude gain at the receiver, thereby reducing the S-BPSK discrimination problem to an ``amplified Kennedy nulling problem''. Furthermore, the PNR detector provides the necessary statistical degrees of freedom for optimal MAP decisions under non-ideal conditions, which will be discussed later.

\begin{figure*}[htbp]
  \centering
  \includegraphics[scale=1.0]{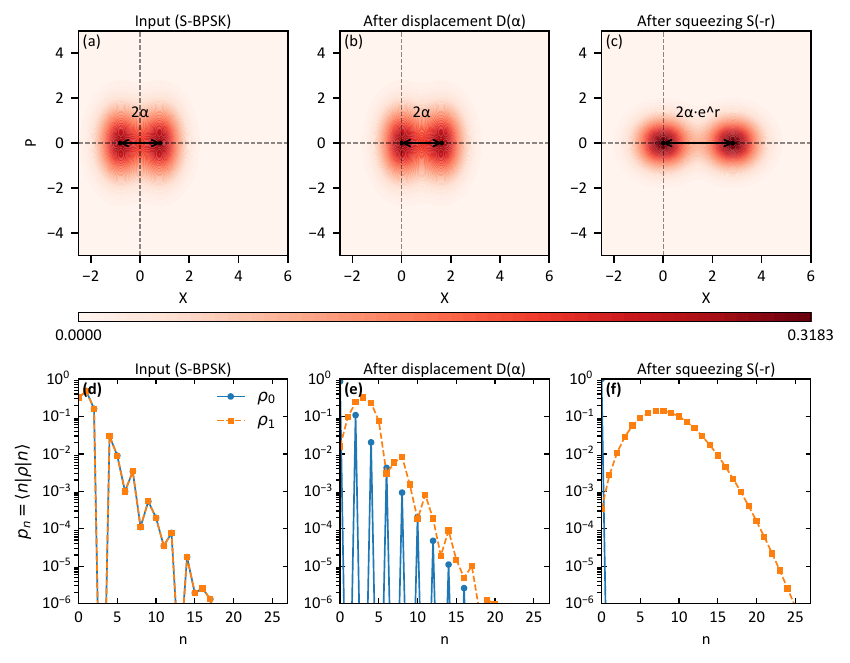}
  \caption{Evolution of received states in the proposed inverse-squeezing Kennedy receiver.
  Panels (a)-(c) show phase-space representations, while panels (d)-(f) show the corresponding Fock-basis populations.
  Panels (a) and (d) depict the received states $\rho_i$.
  Panels (b) and (e) display the states after the nulling displacement $\sigma_i=D(\alpha)\rho_i D^\dagger(\alpha)$.
  Panels (c) and (f) present the states after inverse-squeezing $\zeta_i=U\rho_i U^\dagger$ with $U=S(-r)D(\alpha)$.
  The mean photon number is $N=1.0$ for all panels.}
  \label{fig:state_change}
\end{figure*}

Let the input S-BPSK signal be denoted as $\rho_i \sim \ket{\psi_i}$, where $\ket{\psi_i} = D(\pm \alpha)S(r)\ket{0}$. First, the input signal interferes with a local oscillator (LO) on a highly transmissive beam splitter ($\tau \to 1$) to perform the Kennedy nulling displacement. Consequently, the S-BPSK alphabet is mapped to a squeezed on-off keying (S-OOK) alphabet $\sigma_i$:
\begin{equation}
  \sigma_i \sim D(\alpha)\ket{\psi_i} = \left\{\begin{aligned}
    &S(r)\ket{0}, & i = 0,\\
    &D(2\alpha)S(r)\ket{0}, & i = 1.
  \end{aligned}\right.
  \label{eq:ISK_sigma_compact}
\end{equation}
Subsequently, $\sigma_i$ passes through the IS module, mapping the S-OOK alphabet to a coherent-state on-off keying (C-OOK) alphabet $\zeta_i$:
\begin{equation}
  \zeta_i = S(-r) \sigma_i S^\dagger(-r) \sim \left\{\begin{aligned}
    &S(-r)S(r)\ket{0} = \ket{0}, & i = 0,\\
    &S(-r)D(2\alpha)S(r)\ket{0} = D(2\alpha \mathrm{e}^r)\ket{0} = \ket{2\gamma}, & i = 1.\\
  \end{aligned} \right.
  \label{eq:ISK_zeta_compact}
\end{equation}
In the general case, $\gamma = \alpha\cosh r + \alpha^{*}\sinh r$. Under the condition $\alpha \in \mathbb{R}_+$ adopted in this work, $\gamma = \alpha e^{r}$. Therefore, we define:
\begin{equation}
  \gamma \triangleq \alpha e^{r},\qquad
  N_{\text{eff}} \triangleq |\gamma|^2 = |\alpha|^2 e^{2r}.
  \label{eq:ISK_Neff_general}
\end{equation}
With the optimal energy allocation $\beta_\text{opt}$, the effective photon number can be further expressed as $N_{\text{eff}}=N(N+1)>N$. This indicates that the IS module transforms the transmitter's squeezing resources into a ``coherent-state separation gain'' directly accessible by the receiver. This results in a larger Euclidean distance between the two states in phase space (Fig.~\ref{fig:state_change}(a-c)) and simultaneously reduces the overlap of the Fock-basis populations of $\zeta_i$ (Fig.~\ref{fig:state_change}(e-f)).

Next, PNR detection is performed on $\zeta_i$, yielding an output photon number $n$. The positive operator-valued measure (POVM) for an ideal PNR detector is given by:
\begin{equation}
  \Pi_n^{\text{ideal}} = \ket{n}\bra{n},\quad n=0,1,2,\dots .
\end{equation}
From Eq.~\eqref{eq:ISK_zeta_compact}, we have $\zeta_0=\ket{0}\bra{0}$ and $\zeta_1=\ket{2\gamma}\bra{2\gamma}$, with mean photon numbers $\mu_0 = 0$ and $\mu_1 = |2\gamma|^2 = 4N_{\text{eff}}$, respectively. Thus, the probability of detecting $n$ photons given transmitted symbol $i$ is:
\begin{equation}
  P(n|i)=\text{Tr}\left(\zeta_i\Pi_n^\text{ideal}\right) = \left\{\begin{aligned}
    &\text{Poiss}\left(n;\mu_0\right) = \delta_{0,n}, & i = 0,\\
    &\text{Poiss}\left(n;\mu_1\right), & i = 1,
  \end{aligned}\right.
  \label{eq:ISK_Pn_i_compact}
\end{equation}
where $\text{Poiss}\left(n;\mu\right) = \mathrm{e}^{-\mu}\frac{\mu^n}{n!}$.

The posterior probabilities are given by:
\begin{equation}
  P(i=0|n) = \frac{\pi_0 P(n|i=0)}{\sum_k \pi_k P(n|i=k)},\quad
  P(i=1|n) = \frac{\pi_1 P(n|i=1)}{\sum_k \pi_k P(n|i=k)}.
\end{equation}
The MAP decision rule is defined as:
\begin{equation}
  \hat{i}(n)=
  \begin{cases}
    1, & P(i=1|n) \ge P(i=0|n),\\
    0, & \text{otherwise}.
  \end{cases}
  \label{eq:ISK_MAP_def}
\end{equation}
where $\hat{i}$ is the decision of IS-Kennedy receiver. 
Assuming equal priors ($\pi_0 = \pi_1 = 1/2$), the log-likelihood ratio is:
\begin{equation}
  \label{eq:likelihood_ideal}
  L(n) = \ln \frac{P(n|i=0)}{P(n|i=1)} = \ln \frac{\text{Poiss}\left(n;\mu_0\right)}{\text{Poiss}\left(n;\mu_1\right)} = -\mu_0+\mu_1 + n\ln \left(\frac{\mu_0}{\mu_1}\right).
\end{equation}
Since $L(n)$ is a monotonic function of $n$, the MAP decision (Eq.~\eqref{eq:ISK_MAP_def}) is equivalent to a threshold detection:
\begin{equation}
  \hat{i}(n)=
  \begin{cases}
    1, & n \ge n_\text{th}^*,\\
    0, & \text{otherwise}.
  \end{cases}
\end{equation}
The optimal threshold $n_\text{th}^*$ corresponds to the zero-crossing of the log-likelihood ratio:
\begin{equation}
  \label{eq:threshold_ideal}
  n_\text{th}^* = \left\lceil \frac{\mu_1 - \mu_0}{\ln\mu_1 - \ln\mu_0} \right\rceil.
\end{equation}
Under ideal conditions where $\mu_0 = 0$, we have $\lim_{\mu_0\to 0}\frac{\mu_1 - \mu_0}{\ln\mu_1 - \ln\mu_0}\to 0^+$; thus, $n_\text{th}^* = 1$. This corresponds to a standard on-off detection strategy, implying that a simple SPD suffices to achieve PNR performance in this ideal condition. 
It is important to emphasize, however, that the introduction of the PNR detector is not intended to improve the performance of the IS-Kennedy under ideal condition. 
Rather, it addresses practical system imperfections: once detector imperfections (efficiency/dark counts) or environmental noise occur, $\zeta_0$ deviates from the vacuum state, yielding a non-zero $P(n\ge 1|i=0)$. 
In such cases, the optimal decision is no longer a simple on-off rule, and the PNR detector provides the necessary statistical information to implement a generalized MAP or optimal threshold strategy. 
Quantitative analysis of these non-ideal factors is presented in subsequent sections.

The false alarm probability $P_\text{FA}$ (deciding 1 when 0 is sent) and the miss detection probability $P_\text{Mi}$ (deciding 0 when 1 is sent) are:
\begin{equation}
  \label{eq:probability_falseAlarm_missDetection}
  P_\text{FA} = P\left(n\ge n_\text{th}^* \mid i=0\right),\quad P_\text{Mi} = P\left(n < n_\text{th}^* \mid i=1\right).
\end{equation}
Therefore, under equal priors and ideal conditions ($n_\text{th}^*=1$), the average error probability of the IS-Kennedy is:
\begin{equation}
  \label{eq:ISK_p_err_ideal}
  \begin{aligned}
    P_{\text{err}}^{\text{ISK},\text{ideal}}
    &= \frac{1}{2} \left(P_\text{FA} + P_\text{Mi}\right) \\
    &= \frac{1}{2}\left[ 1 - P(n=0|i=0) + P(n=0|i=1) \right] \\
    &= \frac{1}{2}\exp\left(-|2\gamma|^2\right) = \frac{1}{2}\exp\left(-4N_{\text{eff}}\right).
  \end{aligned}
\end{equation}
Substituting the optimal energy allocation condition $N_{\text{eff}}=N(N+1)$, we obtain:
\begin{equation}
  P_{\text{err}}^{\text{ISK},\text{ideal}}
  = \frac{1}{2}\exp\left[-4N(N+1)\right].
  \label{eq:ISK_p_err_N_compact}
\end{equation}

Notably, we observe that:
\begin{equation}
  \label{eq:ISK_K_relationship}
  P_{\text{err}}^{\text{ISK},\text{ideal}}(N)
  = P_{\text{err}}^{\text{K},\text{ideal}}(N_{\text{eff}}),
\end{equation}
where $P_{\text{err}}^{\text{K},\text{ideal}}$ is the error probability of the conventional Kennedy receiver under ideal conditions~\cite{Kennedy_Receiver}. This implies that, under matched squeezing, the IS-Kennedy is equivalent to mapping an S-BPSK signal with mean photon number $N$ to a C-BPSK signal with mean photon number $N_{\text{eff}}$ via the IS module, followed by Kennedy nulling and on-off detection. That is, the IS-Kennedy converts the squeezing resources of S-BPSK signals into higher coherent-state energy advantage ($N \to N_\text{eff} = N(N + 1)$) through the IS module, thereby achieving a lower error probability. Compared to quantum receivers for coherent state discrimination, the energy gain brought by the IS-Kennedy is
\begin{equation}
  G = 10\cdot \log_{10} \left(\frac{N_\text{eff}}{N}\right) 
  = 10\cdot \log_{10} \left(N+1\right)\quad (\text{dB})
\end{equation}

\begin{figure*}[htbp]
  \centering
  \includegraphics[scale=1.0]{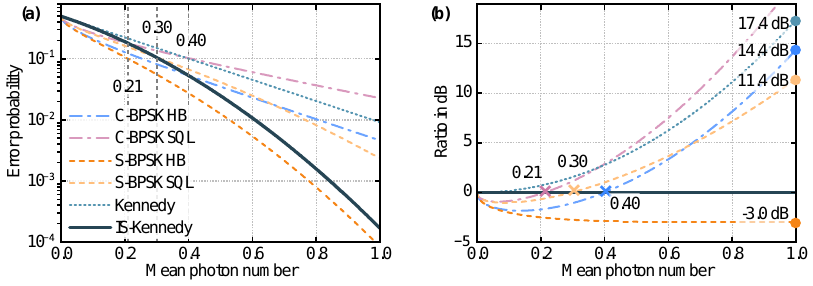}
  \caption{Performance of the proposed inverse-squeezing Kennedy receiver (IS-Kennedy) under ideal conditions. (a) Error probability of the IS-Kennedy ($P_\text{err}^\text{ISK}$). (b) Ratio (in dB) of the error probability limits (Helstrom bound (HB) and standard quantum limit (SQL)) of binary phase-shift keyed displaced squeezed states (S-BPSK) and coherent states (C-BPSK) to $P_\text{err}^\text{ISK}$ (as the benchmark).}
  \label{fig:performance_Ideal}
\end{figure*}

For S-BPSK, the Helstrom bound in Eq.~\eqref{eq:DSS_HB_opt} and the IS-Kennedy error probability in Eq.~\eqref{eq:ISK_p_err_N_compact} satisfy the relation:
\begin{equation}
  \frac{P_\text{err}^{\text{ISK},\text{ideal}}}{P_\text{HB}^\text{S-BPSK}} 
  = \frac{\exp\left\{-4N\left(N+1\right)\right\}}{1 - \sqrt{1 - \exp\left\{-4N(N+1)\right\}}}.
\end{equation}
In the low-energy limit $N\to 0$, both $P_\text{err}^\text{ISK}$ and $P_\text{HB}^\text{S-BPSK}$ approach $0.5$, and their ratio approaches $1$, corresponding to random guessing. In the high-energy regime $N\gg 1$, the ratio approaches $2$ (i.e., $3$~dB), yielding:
\begin{equation}
  P_\text{err}^{\text{ISK},\text{ideal}} \simeq 2 P_\text{HB}^\text{S-BPSK}.
\end{equation}
Although the IS-Kennedy remains suboptimal with respect to the S-BPSK Helstrom bound, satisfying $P_\text{err}^{\text{ISK}} \in \left[P_\text{HB}^\text{S-BPSK}, 2P_\text{HB}^\text{S-BPSK}\right]$ for all $N$, it provides a substantial advantage over C-BPSK channels. Specifically, for the same mean photon number $N$, the proposed IS-Kennedy achieves an error probability strictly lower than the Helstrom bound $P_\text{HB}^\text{C-BPSK}$ of a C-BPSK channel. For instance, around $N\approx 0.4$ in Fig.~\ref{fig:performance_Ideal}, the IS-Kennedy performance lies between the S-BPSK Helstrom bound and the C-BPSK Helstrom bound.

Figure~\ref{fig:performance_Ideal}(a) shows the IS-Kennedy error probability as a function of the mean photon number $N$ under ideal conditions, alongside the Kennedy receiver and benchmarks for C-BPSK and S-BPSK discrimination. The IS-Kennedy surpasses the C-BPSK SQL, the S-BPSK SQL, and the C-BPSK Helstrom bound at approximately $N \approx 0.21$, $0.30$, and $0.40$, respectively. The performance advantage over these benchmarks increases with $N$, consistent with the fact that $P_{\text{err}}^{\text{ISK}}=\frac{1}{2}\exp[-4N(N+1)]$ features an approximately quadratic exponent $N(N+1)$, leading to a steeper decay than the linear scaling of coherent-state benchmarks. The dB ratios summarized in Fig.~\ref{fig:performance_Ideal}(b) quantify these advantages. At $N=1.0$, the IS-Kennedy achieves gains of $21.3$~dB, $11.4$~dB, and $14.4$~dB relative to the C-BPSK SQL, the S-BPSK SQL, and the C-BPSK Helstrom bound, respectively. Additionally, the IS-Kennedy significantly outperforms the conventional Kennedy receiver, exhibiting a $17.4$~dB gain at $N=1.0$. Meanwhile, it remains $3$~dB above the S-BPSK Helstrom bound, staying within a constant factor of the ultimate quantum limit. These results demonstrate that the IS-Kennedy delivers robust performance over a broad photon-number regime under ideal conditions.

The above analysis assumes ideal, noiseless conditions. In the next section, we examine the impact of non-ideal factors on receiver performance, including imperfect photon detection, phase diffusion, and thermal noise.

\subsection*{Imperfect PNR detector with detection efficiency, dark count, and finite resolution}
The previous section presented the closed-form error probability of the IS-Kennedy receiver under ideal detection conditions, pointing out that with perfect nulling and matched squeezing, the optimal decision rule degenerates to a simple on--off strategy. However, in practical implementations, PNR detectors possess finite resolution (maximum resolvable photon number, M) and suffer from non-ideal factors such as limited detection efficiency $\eta$ and dark counts $\nu$ (background counts). In this section, while maintaining the assumption of an ideal IS-Kennedy processing module (i.e., retaining $\zeta_0=\ket{0}\bra{0}$ and $\zeta_1=\ket{2\gamma}\bra{2\gamma}$), we analyze the impact of imperfect PNR detection on the error probability and derive the expression for the optimal threshold (MAP-equivalent) decision and the corresponding error probability.

Considering an imperfect detection efficiency $\eta \in (0,1]$ and dark counts $\nu \ge 0$, the POVM operators are described by~\cite{POVM_Imperfect,dark_count}:
\begin{equation}
  \Pi_n^{\left(\eta,\nu\right)}=\mathrm{e}^{-\nu} \sum_{l=0}^{n} \sum_{k=n-l}^{\infty} \frac{\nu^l}{l!} \binom{k}{n-l}\eta^{n-l}(1-\eta)^{k-(n-l)}\ket{k}\bra{k}.
\end{equation}
Taking the PNR detector's resolution M into account (denoted as PNR(M)), the POVM of a PNR(M) detector is
\begin{equation}
  \Pi_n^{\left(\eta,\nu,\text{M}\right)} = \left\{\begin{aligned}
    & \Pi_n^{\left(\eta,\nu\right)},\quad n = 0,1,\cdots,\text{M}-1,\\
    & \sum_{m=M}^{\infty} \Pi_m^{\left(\eta,\nu\right)},\quad n = \text{M}.
  \end{aligned}\right.
\end{equation}
Consequently, the probability of detecting $n$ photons (Eq.~\eqref{eq:ISK_Pn_i_compact}) is reformulated as:
\begin{equation}
  P(n|i)=\text{Tr}\left\{\zeta_i\Pi_n^{\left(\eta,\nu,\text{M}\right)}\right\} = \left\{\begin{aligned}
    &\text{Poiss}\left(n;\eta\mu_i + \nu\right),\quad n \le \text{M}-1,\\
    &\sum_{m=M}^{\infty}\text{Poiss}\left(m;\eta\mu_i + \nu\right),\quad n = \text{M},
  \end{aligned}\right.
\end{equation}
where $\mu_0 = 0$ and $\mu_1 = \left|2\gamma\right|^2 = 4N_\text{eff}$.
The log-likelihood ratio (Eq.~\eqref{eq:likelihood_ideal}) is updated to:
\begin{equation}
  L(n) = -\eta\mu_0 + \eta\mu_1 + n \ln\frac{\eta\mu_0 + \nu}{\eta\mu_1 + \nu},
\end{equation}
which remains a monotonic function of $n$. Therefore, the MAP criterion is equivalent to a threshold detection scheme. Therefore, the optimal threshold (Eq.~\eqref{eq:threshold_ideal}) involves a truncation at M:
\begin{equation}
  n_\text{th}^* = \min \left\{ \left\lceil \frac{\eta\mu_1 - \eta\mu_0}{\ln\left(\eta\mu_1 + \nu\right) - \ln\left(\eta\mu_0 + \nu\right)} \right\rceil, \text{M} \right\} 
  = \min \left\{ \left\lceil \frac{4\eta N_\text{eff}}{\ln\left(4\eta N_\text{eff} + \nu\right) - \ln\nu} \right\rceil, \text{M} \right\}.
  \label{eq:threshold_imperfect_PNR}
\end{equation}

From the above equation, the adaptive behavior of the IS-Kennedy receiver against different non-ideal factors can be observed:
\begin{itemize}
  \item {Presence of dark counts ($\nu > 0$):} The denominator decreases, resulting in an optimal threshold $n_\text{th}^* > 1$. This implies that the PNR detector can filter out background noise by raising the decision threshold, demonstrating its robustness advantage over a standard SPD.
  \item {Absence of dark counts ($\nu = 0$):} The denominator approaches infinity, leading to $n_\text{th}^* \to 1$. In this limit, the optimal strategy reverts to the standard on--off decision. This indicates that in a pure loss channel ($\eta < 1, \nu=0$), the PNR detector offers no performance gain over an SPD, as simple presence/absence detection is optimal.
\end{itemize}

\begin{figure*}[htbp]
  \centering
  \includegraphics[scale=1.0]{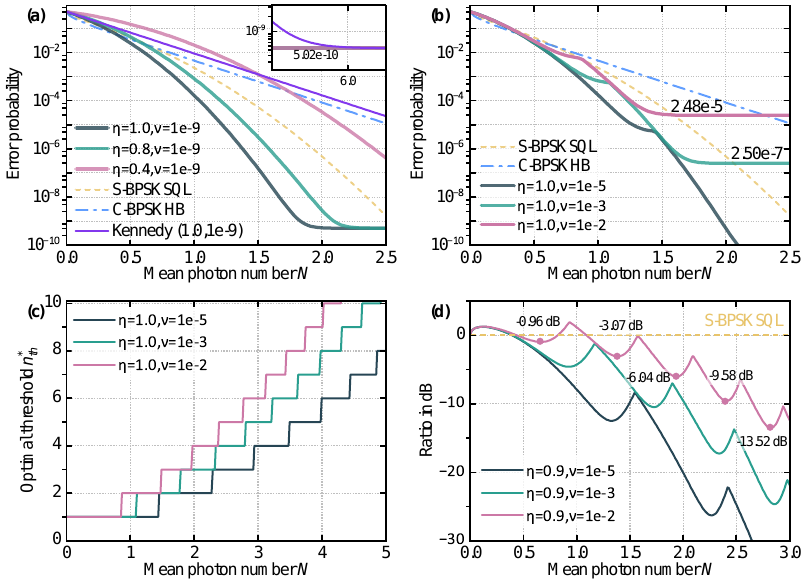}
  \caption{Performance of the proposed inverse-squeezing Kennedy receiver (IS-Kennedy) using an imperfect photon-number-resolution (PNR) detection with finite photon number resolution M (PNR(M)). 
  (a) Error probability of IS-Kennedy with PNR(1) (i.e. single-photon detector) for different detection efficiencies $\eta \in \left\{1.0,0.8,0.4\right\}$ and extremely small dark counts $\nu = 10^{-9}$ (thick solid); the error probability of Kennedy receiver with $\eta=1,\ \nu=10^{-9}$ (solid purple); the Helstrom bound (HB) of binary phase-shift keyed (BPSK) coherent states (C-BPSK) (dot-dash blue); and the standard quantum limit (SQL) of BPSK squeezed states (S-BPSK) (short-dash yellow).
  (b) Error probability of IS-Kennedy with PNR(2) for different dark counts $\nu\in \left\{10^{-2}, 10^{-3},10^{-5}\right\}$ and ideal detection efficiency $\eta=1$ (thick solid).
  (c) The optimal detection threshold $n_\text{th}^*$ as a function of signal energy for different dark counts $\nu\in \left\{10^{-2}, 10^{-3},10^{-5}\right\}$ (solid). The jumps in threshold correspond exactly to the inflection points observed in panels (b) and (d).
  (d) The ratio (in dB) of the IS-Kennedy with PNR(10) (solid) to the S-BPSK SQL (as the benchmark, short-dash yellow). The oscillatory behavior reflects the discrete adjustment of $n_\text{th}^*$ to balance signal detection against dark count noise.}
  \label{fig:performance_nonideal_PNR}
\end{figure*}

The parameter ranges considered in Fig.~\ref{fig:performance_nonideal_PNR} are chosen to represent experimentally relevant classes of photon-number-resolving detectors. Near-unity efficiencies with negligible dark counts model state-of-the-art superconducting PNR detectors, such as segmented superconducting nanowire devices, for which system detection efficiencies approaching 98\% with low dark-count rates have recently been reported in Ref.~\cite{PNR_futong}. 
By contrast, a markedly lower-efficiency regime, around $\eta \approx 0.4$, is representative of silicon photomultiplier (SiPM) or multi-pixel photon-counter platforms, which have been successfully characterized and exploited in quantum-optics experiments despite their lower nominal efficiency and stronger spurious effects such as dark counts and optical cross-talk \cite{R2_01,R2_02}. 
To address this experimentally relevant drastic scenario, we include an additional low-efficiency curve ($\eta = 0.4$) in Fig.~\ref{fig:performance_nonideal_PNR}(a).

As shown in Fig.~\ref{fig:performance_nonideal_PNR}(b) and (d), the error probability and the gain (in dB) relative to the S-BPSK SQL exhibit a characteristic oscillatory or ``step-like'' behavior when dark counts are significant. This phenomenon is intrinsic to the discreteness of PNR detection. As the signal energy $N$ increases, the optimal likelihood ratio varies continuously, whereas the physical decision threshold $n_\text{th}^*$ must remain an integer. Consequently, $n_\text{th}^*$ stays constant over intervals of $N$ (see Fig.~\ref{fig:performance_nonideal_PNR}(c)) before discretely jumping to the next integer level. These jumps allow the receiver to adaptively filter out higher dark-count rates, maintaining robust performance even when $\nu$ is large (e.g., $\nu = 10^{-2}$).

Notably, Fig.~\ref{fig:performance_nonideal_PNR}(a) reveals that while the detection efficiency $\eta$ impacts the rate of error decay (the slope), the saturation error floor at high energies is dictated solely by the dark count rate $\nu$.
In the high-energy regime ($N \gg 1$), the miss detection probability (Eq.~\eqref{eq:probability_falseAlarm_missDetection}) rapidly approaches zero. Consequently, the error probability (saturation floor) is dominated by the false alarm probability, with the optimal threshold saturating at $n_\text{th}^* = \text{M}$:
\begin{equation}
  \begin{aligned}
    P_\text{sat} \approx \frac{1}{2} P_\text{FA} 
    = \frac{1}{2} \mathrm{e}^{-\nu}\sum_{n=\text{M}}^{+\infty}\frac{\nu^n}{n!}
    \xrightarrow{\text{Taylor series}} \frac{1}{2} \left(1 - \nu + \cdots\right)\sum_{n=\text{M}}^{+\infty}\frac{\nu^n}{n!}.
  \end{aligned}
\end{equation}
Given that the dark count rate of practical PNR detectors is typically much less than unity ($\nu \ll 1$)~\cite{PNR_futong}, the dominant contribution to the summation comes from the first term, $n = \text{M}$. Thus, the saturation error probability can be approximated as:
\begin{equation}
  P_\text{sat} \approx \frac{1}{2} \frac{\nu^{\text{M}}}{\text{M}!}.
\end{equation}
This expression indicates that the saturation error probability of the IS-Kennedy receiver is independent of detection efficiency $\eta$ and mean photon number $N$, depending instead on the PNR detector's resolution M and dark count rate $\nu$. Therefore, the existence of any non-zero dark count, however small, inevitably leads to an error probability saturation.
Furthermore, in conjunction with Eq.~\eqref{eq:ISK_K_relationship}, since the advantage of the IS-Kennedy equipped with an SPD lies in converting squeezing resources into effective energy gain rather than suppressing noise floor, its saturation error probability is identical to that of the conventional Kennedy receiver (see Fig.~\ref{fig:performance_nonideal_PNR}(a)).

\subsection*{Phase Diffusion Noise in communication channel}
\begin{figure}[htbp]
  \centering
  \includegraphics[scale=1.0]{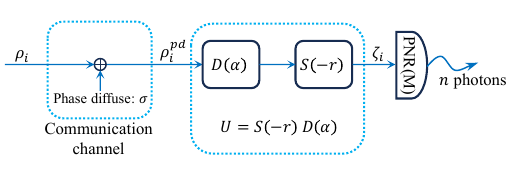}
  \caption{Schematic of the proposed inverse-squeezing Kennedy receiver (IS-Kennedy) under phase diffusion channel. The signal state $\rho_i$ (pure state) is transformed into the state $\rho_i^\text{pd}$ (mixed state) in the phase diffusion channel.}
  \label{fig:structure_phasediffuse}
\end{figure}

Phase diffusion noise is a common communication channel noise that increases the overlap between states, making them more difficult to distinguish \cite{Olivares_03, Olivares2018, phase_diffuse_2019}. A schematic of a phase diffusion noise channel is shown in Fig.~\ref{fig:structure_phasediffuse}.

\begin{figure*}[htbp]
  \centering
  \includegraphics[scale=1.0]{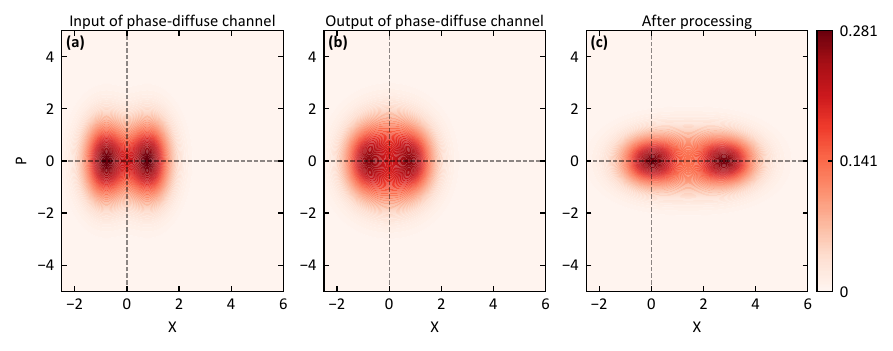}
  \caption{The evolution of phase-space representations of binary phase-shift keyed (BPSK) displaced squeezed states (S-BPSK). (a) S-BPSK $\rho_i$ without phase diffusion noise; (b) states $\rho_i^{\text{pd}}$ with phase diffusion noise; (c) states $\zeta_i$ after the processing of the proposed receiver.
  Parameters for the panels: mean photon number $N=1.0$ and phase diffusion strength $\sigma = 0.5$.}
  \label{fig:phase_space_phase_diffuse}
\end{figure*}

After S-BPSK signals are contaminated by phase diffusion noise in the channel, they evolve from pure states to mixed states and can only be expressed using density operators. Let the uncontaminated signal be denoted as $\rho_i=\left|\psi_i\right\rangle\left\langle\psi_i\right|$. The expression for the contaminated quantum state $\rho_i^{\text{pd}}$ is then:
\begin{equation}
  \label{eq:BPSK_DSS_phaseDiffuse}
  \rho_i^{\text{pd}} =\int_{-\infty}^{+\infty} \mathrm{d}\phi \cdot \left\{ g_{\sigma}(\phi)\cdot \rho_{i,\phi}\right\}
  =\int_{-\infty}^{+\infty} \mathrm{d}\phi \cdot \left\{ g_{\sigma}(\phi)\cdot R_{\phi}\rho_i R_{\phi}^\dagger \right\}    
\end{equation}
Here, $g_\sigma(\phi)=\frac{1}{\sqrt{2\pi\sigma^2}}\exp\left\{ -\frac{\phi^2}{2\sigma^2}\right\}$ is a Gaussian function with mean 0 and standard deviation $\sigma$, which quantifies the noise strength. The rotation operator $R_{\phi}=\mathrm{e}^{-j\phi a^\dagger a}$ rigidly rotates the entire Wigner function distribution of the quantum state by an angle $\phi$ in phase space. We denote the rotated quantum state as 
\begin{equation}
  \label{eq:states_phaseDiffuse_psi}
  \rho_{i,\phi} \sim \left|\psi_{i,\phi}\right\rangle =R_{\phi} \left|\psi_i\right\rangle= D\left( (-1)^{i+1}\alpha \mathrm{e}^{-j\phi} \right)S\left( r \mathrm{e}^{-j2\phi} \right) \left| 0 \right\rangle
\end{equation} 
Thus, phase diffusion noise applies a Gaussian-distributed random phase rotation to the input signal, resulting in a completely positive map that transforms the input Gaussian state $\rho_i$ into a non-Gaussian state $\rho_i^{\text{pd}}$. The phase-space evolution of quantum state $\rho_i$ under phase diffusion noise is illustrated in Fig.~\ref{fig:phase_space_phase_diffuse}, with signal mean photon number $N=1.0$ and a noise strength of $\sigma=0.5$.

For the IS-Kennedy receiver, since the input state $\rho_i^{\text{pd}}$ is no longer Gaussian, the quantum state $\zeta_i$ after displacement-squeezing processing and prior to PNR detection also ceases to be Gaussian.
\begin{equation}
  \begin{aligned}
    \zeta_i &= U\rho_i^{\text{pd}}U^\dagger = \int_{-\infty}^{+\infty} \mathrm{d}\phi \cdot \left\{ g(\phi)\cdot UR_{\phi}\rho_i R_{\phi}^\dagger U^\dagger \right\} \\
    &= \int_{-\infty}^{+\infty} \mathrm{d}\phi \cdot \left\{ g(\phi)\cdot \left| \psi_{i,\phi}' \right\rangle \left\langle \psi_{i,\phi}' \right| \right\}    
  \end{aligned}
\end{equation}
where, 
\begin{equation}
  \left| \psi_{i,\phi}' \right\rangle = U \cdot \left| \psi_{i,\phi} \right\rangle = S\left( -r \right) D\left( (-1)^{i+1}\alpha \mathrm{e}^{-j\phi} + \alpha\right)S\left( r \mathrm{e}^{-j2\phi} \right) \left| 0 \right\rangle 
\end{equation}

Therefore, when measuring the quantum state $\zeta_i$ with PNR(M), the probability of obtaining a detection result of $n$ photons is:
\begin{equation}
  P(n|i) = \text{Tr}\left\{ \Pi_n^{(\text{M})} \cdot \zeta_i \right\} = \int_{-\infty}^{+\infty} \mathrm{d}\phi\cdot g_\sigma(\phi) \cdot \text{Tr}\left\{ \Pi_n^{(\text{M})} \cdot \left| \psi_{i,\phi}' \right\rangle \left\langle \psi_{i,\phi}' \right| \right\},
  \label{eq:n_prob_df}
\end{equation}
where 
\begin{equation}
  \Pi_n^{\left(\text{M}\right)} = \left\{\begin{aligned}
    & \Pi_n^\text{ideal},\quad n = 0,1,\cdots,\text{M}-1,\\
    & \sum_{m=M}^{\infty} \Pi_m^\text{ideal},\quad n = \text{M}.
  \end{aligned}\right.
\end{equation}
By substituting Eq.~\eqref{eq:n_prob_df} into the MAP rule in Eq.~\eqref{eq:ISK_MAP_def}, we numerically determine the optimal decision rule and the corresponding error probability under phase diffusion noise. The optimal threshold is obtained numerically as described in Sec.~\emph{Methods}.

\begin{figure*}[htbp]
  \centering
  \includegraphics[scale=1.0]{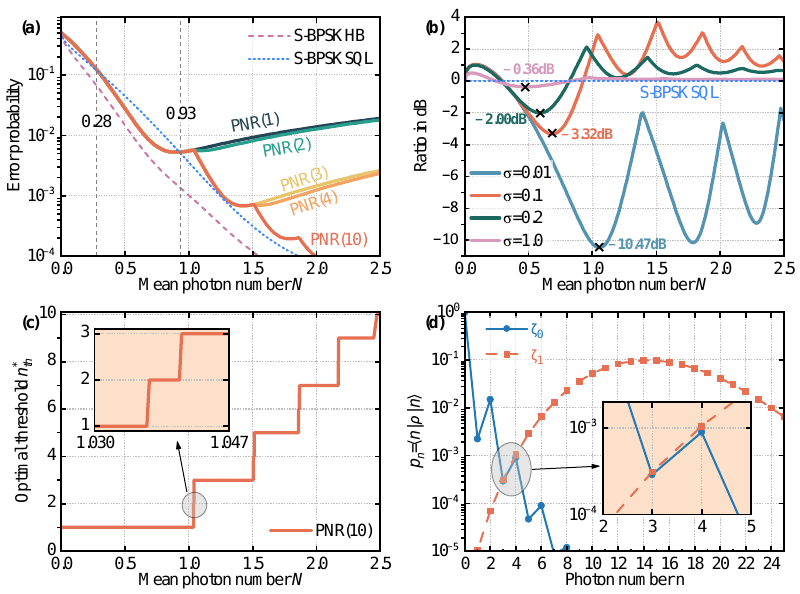}
  \caption{Simulation results of the proposed inverse-squeezed Kennedy (IS-Kennedy) using photon-number-resolving (PNR) detection with finite photon number resolution M (PNR(M)) under phase diffusion noise.
  (a) Error probability of IS-Kennedy for different resolution $\text{M}\in \{1,2,3,4,10\}$ (thick solid), the Helstrom bound (HB, short-dash brown) and standard quantum limit (SQL, short-dot blue) of binary phase-shift keyed (BPSK) displaced squeezed states(S-BPSK). 
  (b) Ratio (in dB) of IS-Kennedy (thick solid) to the S-BPSK SQL (as the benchmark, short-dot blue) for different phase diffusion strength $\sigma \in \{0.01,0.1,0.2,1.0\}$, 
  (c) Optimal threshold $n_{th}^*$ for a resolution of 10,
  and (d) Fock-basis population of measured states $\zeta_i = U \rho_i^\text{pd} U^\dagger$ for $N=1.5$. The analytical expressions for the SQL and HB of S-BPSK under phase diffusion are provided in \emph{Supplementary Information II}.
  Parameters for the panels (a) (c) and (d): phase diffusion strength $\sigma = 0.5$.}
  \label{fig:performance_phasediffuse}
\end{figure*}

Fig.~\ref{fig:performance_phasediffuse}(c) shows that the optimal threshold $n_{\text{th}}^{*}$ under phase diffusion ($\sigma = 0.1$) exhibits a staircase-like dependence on the mean photon number $N$. In the absence of PNR resolution limits, $n_{\text{th}}^{*}$ increases predominantly in unit steps, while some even thresholds remain optimal only over very narrow intervals of $N$, leading to unequal step widths compared with the dark-count case in Sec.~\emph{Results: Imperfect PNR detector with detection efficiency, dark count, and finite resolution}.
This behavior originates from the non-Gaussianity induced by phase diffusion: after the IS-Kennedy processing, $\zeta_i$ cannot be perfectly nulled to (near-)vacuum and its photon-number distribution exhibits noticeable even-odd modulation (Fig.~\ref{fig:performance_phasediffuse}(d)). As $N$ varies, this modulation causes the likelihood function $L\left(n\right)$ to shift non-uniformly across adjacent photon numbers, so that certain even thresholds can be optimal only over short $N$ intervals. Consequently, $n_\text{th}^*$ still changes mainly by +1, but with uneven plateau lengths.
We verified this by refining the sampling of $N$, which reveals short-lived plateaus at some even $n_{\text{th}}^{*}$ values that may be missed under coarser sampling.

Fig.~\ref{fig:performance_phasediffuse}(a) depicts the error probability of IS-Kennedy as a function of the mean signal photon number under a phase diffusion strength of $\sigma = 0.1$. 
And Fig.~\ref{fig:performance_phasediffuse}(b) shows the ratio (in dB) relative to the S-BPSK SQL under $\sigma = 0.01,0.1,0.2,1.0$. 
The results show that IS-Kennedy can surpass the SQL of the S-BPSK scheme within a certain photon-number range ($N \in \left(0.28, 0.93\right)$ under $\sigma = 0.1$), achieving a maximum advantage of 10.47 dB ($\sigma = 0.01$), 3.32 dB ($\sigma = 0.1$), 2.00 dB ($\sigma = 0.2$) and 0.36 dB ($\sigma = 1.0$) beyond the S-BPSK SQL. 

However, at high photon numbers, IS-Kennedy no longer exceeds the SQL. In particular, when the optimal threshold reaches the maximum resolution limit of the PNR detector, the error probability does not saturate but instead increases with the mean photon number. Furthermore, we observe that a PNR detector with a resolution of 2n (even) provides only marginal improvement in IS-Kennedy performance compared to one with a resolution of 2n-1 (odd); no pronounced oscillatory drop in error probability is evident. This behavior is consistent with the variation in the optimal threshold, as shown in panel (c).

\subsection*{Thermal noise in the receiver}
After contamination by thermal noise, the optical states are generally mixed and should be described by density operators. 
Following the Glauber--Sudarshan representation, the impact of thermal background at the receiver can be modeled as a classical, Gaussian-distributed random displacement in phase space, whose statistics are independent of the input state \cite{zhaomufei_2020, zhaomufei_2021_01, zhaomufei_2021_02, Thermal_noise_01}. 
Accordingly, for an arbitrary input state $\rho_i$, the thermal-noise channel can be written as
\begin{equation}
\label{eq:thermal_channel}
  \rho_i^{\mathrm{th}}=\Phi_{n_t}(\rho_i)
  =\int_{\mathbb{C}} \mathrm{d}^2\lambda \, f_{n_t}(\lambda)\, D(\lambda)\rho_i D^\dagger(\lambda),
\end{equation}
where $D(\lambda)=\exp(\lambda a^\dagger-\lambda^\ast a)$ denotes the displacement operator and 
\begin{equation}
  \label{eq:thermal_prob_func}
  f_{n_t}(\lambda)=\frac{1}{\pi n_t}\exp\!\left(-\frac{|\lambda|^2}{n_t}\right)
\end{equation}
is the circularly symmetric complex Gaussian distribution. 
Here $n_t$ is the mean photon number of the thermal background mode, which under thermal equilibrium is
$n_t=\left[\exp\!\left(\frac{h\nu}{kT_0}\right)-1\right]^{-1}$
with $h$ the Planck constant, $k$ the Boltzmann constant, $\nu$ the optical frequency, and $T_0$ the receiver temperature. 
In practical fiber-optic communication systems, the receiver-originated background often dominates over the channel-induced thermal noise \cite{Thermal_noise_02}.

\begin{figure}[htbp]
  \centering
  \includegraphics[scale=1.0]{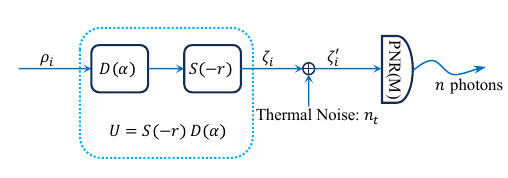}
  \caption{Schematic of the proposed inverse-squeezing Kennedy receiver (IS-Kennedy) under receiver thermal noise. The measured states $\zeta_i$ impinging on the photon-number-resolving (PNR) detection with finite photon number resolution M (PNR(M)) are contaminated by receiver thermal noise and become $\zeta_i'$.}
  \label{fig:Structure_of_ISK_thermal}
\end{figure}

Consistent with Ref.~\cite{zhaomufei_2021_01, zhaomufei_2021_02}, we assume that the thermal-noise contamination occurs immediately before the PNR detector, as illustrated in Fig.~\ref{fig:Structure_of_ISK_thermal}. Therefore the HB and SQL of received states $\rho_i$ remain unchanged.

Consequently, the states $\zeta_i$ measured by PNR(M), originally $\ket{0}$ and $\ket{2\gamma}$, are transformed into
\begin{equation}
\label{eq:zeta_thermal_mixture}
  \begin{aligned}
    \zeta_0' &= \Phi_{n_t}\!\big(|0\rangle\langle 0|\big)
    = \int_{\mathbb{C}}\mathrm{d}^2 \lambda \, f_{n_t}(\lambda)\, \left|\lambda\right\rangle \left\langle \lambda\right|, \\
    \zeta_1' &= \Phi_{n_t}\!\big(\ket{2\gamma}\bra{2\gamma}\big)
    = \int_{\mathbb{C}}\mathrm{d}^2 \lambda \, f_{n_t}(\lambda)\, \left|\lambda + 2\gamma\right\rangle \left\langle \lambda +2\gamma\right|.
  \end{aligned}
\end{equation}
Because the thermal-noise channel $\Phi_{n_t}$ in Eq.~\eqref{eq:thermal_prob_func} is displacement covariant (namely, applying a displacement before the channel is equivalent to applying the same displacement after the channel), Eq.~\eqref{eq:zeta_thermal_mixture} may write 
\begin{equation}
  \zeta_0'=\rho_{\mathrm{th}}(n_t),\quad \zeta_1'=D(2\gamma)\rho_{\mathrm{th}}(n_t)D^\dagger(2\gamma).
  \label{eq:zeta_thermal_mixture_02}
\end{equation}

Thus, under receiver thermal noise, $\zeta_0'$ and $\zeta_1'$ are classical mixtures of coherent states. 
The probability of detecting $n$ photons with PNR(M) is
\begin{equation}
  \label{eq:pnr_prob_thermal}
  P\left(n|i\right)
  =\text{Tr}\!\left( \Pi_n^{\left(\text{M}\right)}\, \zeta_i' \right)
  =\int_{\mathbb{C}} \mathrm{d}^2\lambda \cdot f_{n_t}(\lambda) \cdot \text{Poiss}\left\{n;\mu_i(\lambda)\right\} \end{equation}
where $\mu_0(\lambda)=|\lambda|^2$, $\mu_1(\lambda)=|\lambda+2\gamma|^2$. 
Substituting Eq.~\eqref{eq:pnr_prob_thermal} into the MAP rule in Eq.~\eqref{eq:ISK_MAP_def}, we numerically obtain the optimal decision rule and the corresponding error probability in the presence of receiver thermal noise. The optimal threshold is obtained numerically as described in Sec.~\emph{Methods}.

\begin{figure*}[htbp]
  \centering
  \includegraphics[scale=1.0]{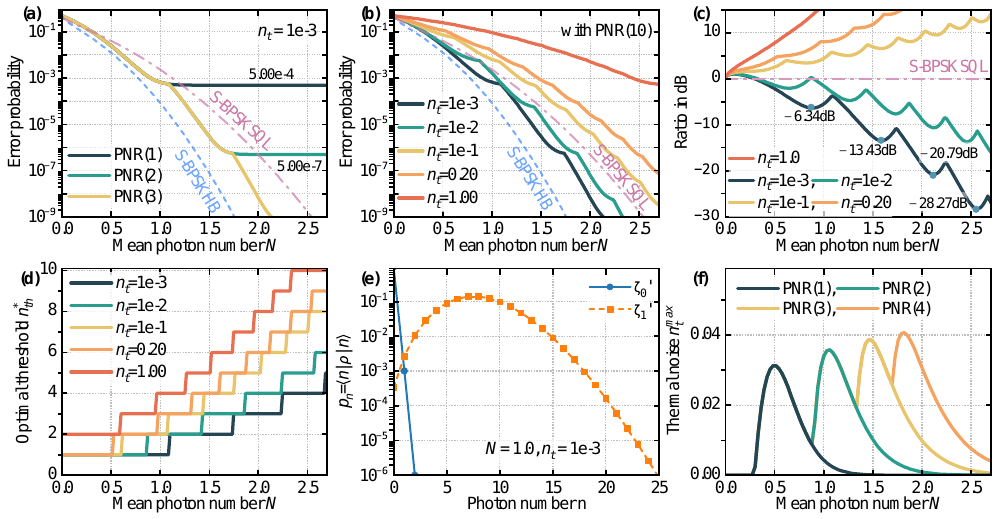}
  \caption{Simulation results of the proposed inverse-squeezing Kennedy receiver (IS-Kennedy) using the photon-number-resolving (PNR) detection with finite photon number resolution M (PNR(M)) under thermal noise condition.
  (a) Error probability of IS-Kennedy for different resolutions $\text{M}\in \{1,2,3\}$ (noise strength $n_t = 0.001$) and the error probability limits (Helstrom bound (HB, short-dash blue) and standard quantum limit (SQL, dot-dash pink)) of binary phase-shift keyed (BPSK) displaced squeezed states (S-BPSK).
  (b) Error probability of the IS-Kennedy ($\text{M}=10$, thick solid) for $n_t \in \{10^{-3},10^{-2},0.1,0.2,1.0\}$. 
  (c) Ratio (in dB) of the IS-Kennedy ($\text{M}=10$, thick solid) to the S-BPSK SQL (as the benchmark, dot-dash pink) for $n_t \in \{10^{-3},10^{-2},0.1,0.2,1.0\}$.
  (d) Optimal threshold of the IS-Kennedy ($\text{M}=10$, thick solid) for $n_t \in \{10^{-3},10^{-2},0.1,0.2,1.0\}$. 
  (e) Fock-basis population of $\zeta_i'$ (the input energy $N = 1.0$ and $n_t = 0.001$), and (f) Maximum tolerable thermal noise $n_t^\text{max}$ for $\text{M}\in \{1,2,3,4\}$.}
  \label{fig:performance_thermal}
\end{figure*}

Fig.~\ref{fig:performance_thermal}(a) and (b) depict the error probability of the IS-Kennedy as a function of the mean signal photon number, under different PNR detector resolutions and various thermal noise strengths. The error probability curves exhibit an oscillatory pattern and decrease overall with increasing signal energy, which is a consequence of the continuous adjustment of the optimal threshold $n_\text{th}^*$ (see panel (d)). This trend continues until $n_\text{th}^*$ reaches the maximum resolution $\text{M}$ of the PNR detector, after which the error probability saturates and no longer varies with the mean photon number. 
As analyzed in Sec.~\emph{Results: Imperfect PNR detector with detection efficiency, dark count, and finite resolution}, the saturated error probability is dominated by the false alarm probability $P_\text{FA}$ (Eq.~\eqref{eq:probability_falseAlarm_missDetection}), and when the signal energy tends to infinity, the optimal threshold must be equal to $\text{M}$. Therefore, the saturated error probability is:
\begin{equation}
  \begin{aligned}
    P_\text{sat} \approx \frac{1}{2}P_\text{FA} &= \frac{1}{2} P\left(n\ge \text{M}|i=0\right) = \frac{1}{2} \sum_{n = \text{M}}^{\infty} \text{Tr}\left\{\ket{n}\bra{n}\cdot \rho_\text{th}\left(n_t\right)\right\}\\
    &= \frac{1}{2}\sum_{n=\text{M}}^{\infty} \frac{{n_t}^n}{(1 + n_t)^{n+1}} = \frac{1}{2}\left(\frac{n_t}{1 + n_t}\right) ^ \text{M}\\
  \end{aligned}
\end{equation}

Fig.~\ref{fig:performance_thermal}(c) shows the ratio (in dB) of the IS-Kennedy error probability to the SQL of S-BPSK versus the mean signal photon number under different noise strengths. It can be observed that the IS-Kennedy can still surpass the SQL at certain noise levels. For instance, when $n_t = 10^{-3}$, the error probability of the IS-Kennedy with a PNR(M) detector is lower than the SQL by -6.34dB (M=1), -13.43dB (M=2), -20.79dB (M=3), and -28.27dB (M=4).

We then introduce the maximum tolerable thermal noise $n_t^\text{max}$ as a figure of merit, defined as the highest noise level for which $P_\text{err}^\text{ISK} \le P_\text{SQL}^\text{S-BPSK}$ at a given signal energy N. This quantity is shown in Fig.~\ref{fig:performance_thermal}(f). Thus, the IS-Kennedy outperforms the SQL when $n_t < n_t^\text{max}$, corresponding to the region below the $n_t^\text{max}$ curve. The plot displays M distinct peaks, after which it decays toward zero.

\subsection*{Transmission loss in the communication channel}
A practically important limitation of squeezed-state communication is transmission loss. 
In contrast to the ideal analysis of Sec.~\emph{Results: inverse-squeezing Kennedy receiver under ideal conditions}, propagation through a lossy channel does not merely reduce the signal displacement, but also degrades the squeezing resource itself by mixing the signal mode with an environmental vacuum mode. 
Equivalently, in the Stinespring representation, a pure-loss bosonic channel can be modeled by a beam splitter of transmittance $T$ that couples the signal to an environmental vacuum state \cite{Olivares2018, review_2012}. 

For nonclassical inputs such as displaced squeezed states, this interaction generally generates system-environment entanglement. However, since the environment is inaccessible to the receiver, the relevant state for discrimination is the reduced output state of the signal mode, obtained after tracing out the environment. Therefore, the received signals are correctly described as single-mode mixed Gaussian state, namely a displaced squeezed thermal state (DSTS), rather than pure displaced squeezed vacuum state \cite{review_2012}.

% \paragraph*{1. Input Gaussian-state description.}
The S-BPSK input alphabet of pure-loss channel is $\ket{\psi_i} = D[(-1)^{i+1}\alpha] S(r) \ket{0}$ with $\alpha, r \in \mathbb{R}_+$. 
As shown in \emph{Supplementary Information I}, each displaced squeezed state is a single-mode Gaussian state fully characterized by its displacement vector and covariance matrix. In the quadrature basis $(\hat X,\hat P)$ one has
\begin{equation}
  \vec{d_i}=\begin{pmatrix}
    (-1)^{i+1}\sqrt{2}\alpha\\
    0
  \end{pmatrix},\qquad
  \mathbf{V}=\frac{1}{2} \begin{pmatrix}
    e^{-2r} & 0\\
    0 & e^{2r}
  \end{pmatrix}.
\end{equation}

\begin{figure*}[htbp]
  \centering
  \includegraphics[scale=1.0]{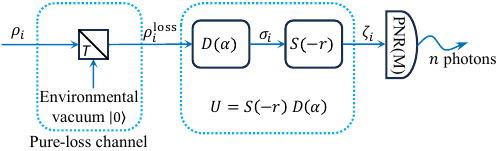}
  \caption{Schematic of the proposed inverse-squeezing Kennedy receiver (IS-Kennedy) with the pure-loss channel. The signal state $\rho_i$ (pure state) is transformed into the state $\rho_i^\text{loss}$ (mixed state) in the pure-loss communication channel, which $T$ is the equivalent transmittance.}
  \label{fig:structure_loss}
\end{figure*}

% \paragraph*{2. Pure-loss channel as a Gaussian map.}
We model propagation by a pure-loss bosonic channel with transmittance $T\in[0,1]$ (see Fig.~\ref{fig:structure_loss}) \cite{review_2012}. At the operator level, the channel can be represented as
\begin{equation}
  \hat a_{\mathrm{out}}=\sqrt{T}\,\hat a_{\mathrm{in}}+\sqrt{1-T}\,\hat v,
\end{equation}
where $\hat v$ is an environmental vacuum mode \cite{review_2012}. 
For displaced squeezed vacuum state, this channel acts linearly on first and second moments:
\begin{equation}
  \label{eq:state_vec_matrix_pureLoss}
  \vec{d_i}^{(T)}=\sqrt{T}\,\vec{d_i},
  \qquad
  \mathbf{V}^{(T)} = T\cdot \mathbf{V} + (1-T)\frac{\mathbf{I}}{2} =
  \frac12\begin{pmatrix}
    1-T+Te^{-2r} & 0\\
    0 & 1-T+Te^{2r}
  \end{pmatrix}.
\end{equation}
Several remarks are immediate: 
(i) When $T=1$, one recovers the ideal pure displaced squeezed states; 
(ii) For $0<T<1$ and $r\neq 0$, the determinant of $\mathbf{V}^{(T)}$ is larger than $1/4$, which means that the reduced output state is mixed.
Therefore, after propagation, the received S-BPSK signals are no longer pure displaced squeezed vacuum states, but displaced squeezed thermal states.
This is precisely the local manifestation of the system--environment coupling discussed above: the information leaked to the inaccessible environment appears at the receiver as a loss-induced mixedness.

% \paragraph*{3. Equivalent squeezed-thermal parametrization.}
It is convenient to rewrite $\mathbf{V}^{(T)}$ in the standard squeezed-thermal form
\begin{equation}
  \mathbf{V}^{(T)}=
  \left(n_T+\frac{1}{2}\right)
  \begin{pmatrix}
    e^{-2r_T} & 0\\
    0 & e^{2r_T}
  \end{pmatrix}.
\end{equation}
By comparing matrix elements, the residual squeezing parameter $r_T$ and the equivalent thermal photon number $n_T$ are obtained as
\begin{equation}
  \begin{aligned}
    r_T&=\frac14\ln\!\left(
      \frac{1-T+Te^{2r}}{1-T+Te^{-2r}}
    \right),\\
    n_T&=\frac12\left[
      \sqrt{(1-T+Te^{-2r})(1-T+Te^{2r})}-1
    \right].
  \end{aligned}
\end{equation}

Therefore, the input of IS-Kennedy is:
\begin{equation}
  \label{eq:rho_pureLoss}
  \rho_i^\text{loss} = \left\{\begin{aligned}
    & D(-\sqrt{T}\alpha) S\left(r_T\right) \rho_\text{th}\left(n_T\right) S^\dagger \left(r_T\right) D^\dagger (-\sqrt{T}\alpha),\quad i = 0\\
    & D(+\sqrt{T}\alpha) S\left(r_T\right) \rho_\text{th}\left(n_T\right) S^\dagger \left(r_T\right) D^\dagger (+\sqrt{T}\alpha),\quad i = 1
  \end{aligned}\right.
\end{equation}
These expressions have a clear physical meaning. 
When $T\to1$, one has $r_T\to r$ and $n_T\to0$, so the ideal case is recovered.
When $0<T<1$, the effective squeezing is reduced from $r$ to $r_T$, while a nonzero thermal component $n_T$ appears.
When $T\to0$, the signal is completely extinguished and both states collapse to vacuum.
Hence, transmission loss affects the receiver in two distinct ways: it attenuates the displacement and, at the same time, destroys the purity of the squeezing resource.

% \paragraph*{4. Matched IS-Kennedy processing after a lossy channel.}
Under ideal conditions, the nulling displacement $D(\alpha)$ and the inverse-squeezing $S(-r)$ are matched to the transmitted state.
After a lossy channel, the received states are centered at $\pm\sqrt{T}\alpha$, rather than at $\pm\alpha$, and exhibit residual squeezing $r_T$ rather than $r$.
Therefore, the matched version of the IS-Kennedy in the lossy scenario should use
\begin{equation}
  D(\alpha)\longrightarrow D(\sqrt{T}\alpha),\quad
  S(-r)\longrightarrow S(-r_T).  
\end{equation}

The nulling displacement $D(\sqrt{T}\alpha)$ maps the attenuated BPSK alphabet into a lossy squeezed OOK alphabet:
\begin{equation}
  \sigma_i = \left\{\begin{aligned}
    & S\left(r_T\right) \rho_\text{th}\left(n_T\right) S^\dagger \left(r_T\right),\quad i = 0\\
    & D(2\sqrt{T}\alpha) S\left(r_T\right) \rho_\text{th}\left(n_T\right) S^\dagger \left(r_T\right) D^\dagger (2\sqrt{T}\alpha),\quad i = 1
  \end{aligned}\right.
\end{equation}
Subsequently, applying the matched inverse-squeezing $S(-r_T)$ removes the residual anisotropy of the covariance matrix. 
As in the ideal case, the inverse-squeezing converts the remaining squeezing resource into a larger effective coherent-state displacement. 
Using the standard transformation of displacement operators under squeezing, one obtains
\begin{equation}
  S(-r_T)D(2\sqrt{T}\alpha)S(r_T)=D(2\sqrt{T}\alpha e^{r_T}).
\end{equation}
Defining
\begin{equation}
  \gamma_T = \sqrt{T}\,\alpha e^{r_T},
\end{equation}
the two output states impinging on the PNR detector become (an OOK pair composed of
a thermal state and a displaced thermal state)
\begin{equation}
  \label{eq:zeta_loss}
  \zeta_i = \left\{\begin{aligned}
    &\rho_{\text{th}}(n_T),\quad i = 0,\\
    &D(2\gamma_T)\rho_{\text{th}}(n_T)D^\dagger(2\gamma_T),\quad i = 1.
  \end{aligned}\right.
\end{equation}
The above formulas show that transmission loss weakens the squeezing advantage through two simultaneous mechanisms.
First, the coherent-state separation gain is reduced from $\gamma=\alpha e^r$ to $\gamma_T=\sqrt{T}\alpha e^{r_T}$.
Second, the perfectly nulled hypothesis is no longer vacuum. Under ideal conditions, the nulling branch produces $\zeta_0=\ket{0}\bra{0}$. After transmission loss, however, $\zeta_0^{(T)}=\rho_{\mathrm{th}}(n_T)$.

Hence, the conditional photon-number probabilities are
\begin{equation}
  P\left(n|i\right)
  =\text{Tr}\!\left( \Pi_n^{\left(\text{M}\right)}\, \zeta_i \right)
  =\int_{\mathbb{C}} \mathrm{d}^2\lambda \cdot f_{n_T}(\lambda) \cdot \text{Poiss}\left\{n;\mu_i(\lambda)\right\}, 
\end{equation}
where $\mu_0(\lambda)=|\lambda|^2$, $\mu_1(\lambda)=|\lambda+2\gamma_T|^2$. 
These expressions are identical in structure to the thermal-noise formulas already used in Sec.~\emph{Results: Thermal noise in the receiver}, 
with the substitutions:
\begin{equation}
  \label{eq:n_T_and_gamma_T}
  n_t \rightarrow n_T,\quad \gamma \rightarrow \gamma_T.
\end{equation}
Since the lossy case reduces to the same thermal-state decision structure with the substitutions in Eq.~\eqref{eq:n_T_and_gamma_T}, the MAP decision rule and the corresponding error probability are evaluated numerically. The optimal threshold is obtained numerically as described in Sec.~\emph{Methods}.

For practical optical communication, one may parameterize the transmittance as
\begin{equation}
  T(a,L)=10^{-aL/10},
\end{equation}
where $a$ is the attenuation coefficient in dB/km and $L$ is the link length in km.
Therefore, once $L$ and $a$ are fixed, the MAP (optimal threshold) and the error probability can be computed numerically without introducing any new statistical machinery.
To model transmission loss in a representative long-haul fiber scenario, we fix the attenuation coefficient at $a = 0.2\ \text{dB/km}$, which is a standard rule-of-thumb value around the 1550 nm telecom window \cite{fiber_loss}.

\begin{figure*}[htbp]
  \centering
  \includegraphics[scale=1.0]{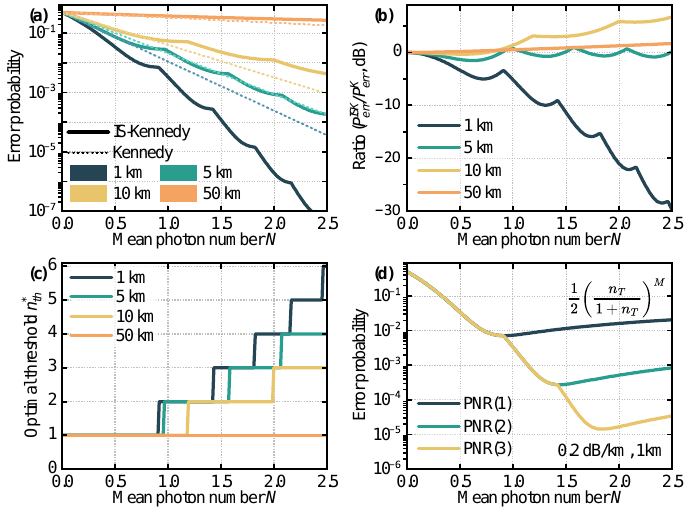}
  \caption{Simulation results of the proposed inverse-squeezing Kennedy receiver (IS-Kennedy) using the photon-number-resolving (PNR) detection with finite photon number resolution M (PNR(M)) under transmission loss condition. 
  (a) Error probabilities of IS-Kennedy (thick solid) and Kennedy receiver (short dash) for different link length $L$ (1 km, 5 km, 10 km, 50 km), 
  (b) Ratio (in dB) of IS-Kennedy (thick solid) to $P_\text{err}^\text{K, loss}$ (the error probability of the lossy Kennedy receiver with matched displacement, $P_\text{err}^\text{K, loss} = \frac{1}{2}e^{-4TN}$, as the benchmark) for different link length $L$ (1 km, 5 km, 10 km, 50 km), 
  (c) Optimal threshold of IS-Kennedy (thick solid) for different link length $L$ (1 km, 5 km, 10 km, 50 km), 
  (d) Error probability of IS-Kennedy (thick solid) for different resolution (M = 1, 2, 3).
  Parameters for the panels: attenuation coefficient $a = 0.2\ \text{dB}/\text{km}$.}
  \label{fig:performance_loss}
\end{figure*}

Figure~\ref{fig:performance_loss}(a) shows that the error probability of the IS-Kennedy increases monotonically with link length. Compared with the lossy Kennedy receiver, the IS-Kennedy curves exhibit a weak oscillatory structure as the input energy increases, whereas the Kennedy curves remain smooth. This difference originates from the threshold adaptation of the IS-Kennedy: as the received signal energy varies, the optimal threshold changes in discrete steps, which produces the visible ripple-like behavior (see Fig.~\ref{fig:performance_loss}(c)). By contrast, the lossy Kennedy receiver retains the standard on-off structure because a coherent-state input remains coherent under pure loss.
A second and more important difference is that, after transmission loss, the nulling branch of the IS-Kennedy is no longer vacuum. For coherent-state signaling, the nulled hypothesis ideally remains vacuum after matched displacement, so the false-alarm probability is determined only by detector imperfections. For squeezed-state signaling, however, the lossy channel transforms the input into a displaced squeezed thermal state; after matched receiver processing, the nulled hypothesis becomes a thermal state rather than vacuum. Consequently, the probability of detecting photons under hypothesis $i=0$ is no longer zero, which explains why loss degrades the IS-Kennedy more strongly than the Kennedy benchmark.

The comparison with the lossy Kennedy receiver in Fig.~\ref{fig:performance_loss}(b) makes this trend more explicit. For low-loss links, the IS-Kennedy retains a clear negative dB ratio over a broad range of $N$, showing that the residual squeezing can still be converted into a receiver-side discrimination gain. As the link length increases, however, the ratio curve shifts upward and eventually approaches or exceeds 0 dB, indicating that the advantage over the Kennedy benchmark is progressively erased by channel attenuation. Therefore, within the fixed source parametrization adopted here, the main benefit of the IS-Kennedy is concentrated in the low-loss regime.

The role of the PNR resolution is illustrated in Fig.~\ref{fig:performance_loss}(d). As analyzed in Sec.~\emph{Results: Thermal noise in the receiver}, when the signal energy increases, the optimal threshold  increases to M, the miss detection probability $P_\text{Mi}$ decreases approaching 0, and the receiver's error probability is dominated by false alarm probability $P_\text{FA}$, so that at very high input energy: 
\begin{equation}
  P_\text{err}^\text{ISK,loss} \to \frac{1}{2}P_\text{FA}(M) 
  = \frac{1}{2}\left(\frac{n_T}{1 + n_T}\right) ^ M
\end{equation}
However, the difference is that $n_T$ varies with the signal energy $N$, so under transmission loss conditions, a saturated error probability does not occur.

\begin{figure}[htbp]
  \centering
  \includegraphics[scale=1.0]{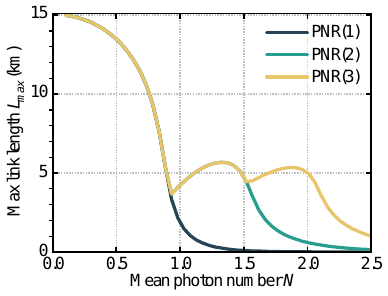}
  \caption{Maximum link length $L_\text{max}$ of the proposed receiver. 
  The benchmark is the error probability of the lossy Kennedy receiver with matched displacement, given by $P_\text{err}^\text{K, loss} = \frac{1}{2}e^{-4TN}$. Results are shown for different mean photon numbers $N$ and finite photon-number resolution $\text{M}\in\{1,2,3\}$. The attenuation coefficient is $a = 0.2\ \text{dB}/\text{km}$.}
  \label{fig:max_length}
\end{figure}

Figure~\ref{fig:max_length} summarizes the practical operating regime through the maximum advantageous link length $L_\text{max}$, defined as the longest link for which the IS-Kennedy still outperforms the lossy Kennedy receiver. Hence, the IS-Kennedy is advantageous in the region $L<L_\text{max}$, i.e., below the corresponding $L_\text{max}$ curve.

Overall, the transmission-loss analysis shows that, within the fixed-$\beta$ S-BPSK parametrization considered here, the IS-Kennedy advantage is concentrated in the low-loss regime.
Since the fixed-$\beta = \frac{N}{2N+1}$ might not remain optimal in a lossy channel, this conclusion should not be interpreted as a general exclusion of long-distance links employing squeezed states.
We anticipate that allowing $\beta$ to adapt with propagation distance would lower the IS-Kennedy error probability and increase the maximum transmission distance $L_\text{max}$, which will be pursued in future work.

\section*{Discussion}
We have proposed and analyzed the inverse-squeezing Kennedy receiver for S-BPSK discrimination. By combining displacement, orthogonally oriented inverse-squeezing, and PNR/MAP detection, the receiver converts transmitter-side squeezing into an effective separation gain at the measurement stage. Under ideal conditions and equal priors, the resulting error probability stays within a constant factor of the S-BPSK Helstrom bound for all N, while surpassing the S-BPSK SQL at $N\approx0.3$ and the C-BPSK Helstrom bound at $N\approx0.4$.

Under nonideal conditions, adaptive thresholding mitigates the impact of detector inefficiency, dark counts, phase diffusion, and receiver thermal noise over practical parameter ranges. Transmission loss is more restrictive: it attenuates the signal and degrades the squeezing resource, thereby reducing the advantage over Kennedy-type and coherent-state benchmarks with increasing propagation distance. Therefore, under the fixed launch-state optimization adopted here, the proposed receiver is most favorable in low-loss channels, while its advantage is progressively reduced as propagation loss increases.

Future work should extend the present analysis beyond the fixed-$\beta$ source parametrization adopted here and consider joint optimization of the launch state and receiver settings in the presence of transmission loss. In particular, re-optimizing the squeezing fraction and nulling displacement as functions of channel attenuation may further improve performance and enlarge the advantageous operating regime of the inverse-squeezing Kennedy receiver.

\section*{Methods}
\subsection*{Threshold determination in non-ideal scenarios}
For the phase-diffusion, receiver-thermal-noise, and transmission-loss analyses, a general closed-form proof of the monotonicity of the likelihood ratio \(L(n)\) is not available. We therefore computed the conditional counting statistics \(P(n|i)\) numerically and evaluated the error probability for all admissible threshold values within the detector-resolution window, \(n_{\mathrm{th}} \in [1,M]\). The threshold was then chosen to minimize the error probability:
\begin{equation}
  P_{\mathrm{err,th}}^{\mathrm{ISK}}=
  \min_{n_{\mathrm{th}} \in [1,M]}
  \frac{1}{2}
  \left[
  P(n<n_{\mathrm{th}} \mid i=1)+
  P(n\ge n_{\mathrm{th}} \mid i=0)
  \right].
\end{equation}

For comparison, we also evaluated the error probability given by the full MAP decision rule:
\begin{equation}
  P_{\mathrm{err,MAP}}^{\mathrm{ISK}}=
  1-\frac{1}{2}\sum_{n=0}^{M}
  \max\!\left\{
  P(n\mid i=0),\, P(n\mid i=1)
  \right\}.
\end{equation}

We found that these two error probabilities coincide numerically over the parameter ranges considered in this work, indicating that, within the explored regimes, the full MAP rule is effectively equivalent to an optimal threshold rule.

\subsection*{Simulation implementation}

All numerical simulations were performed in Python using QuTiP (version: 5.1.1) \cite{qutip}, together with NumPy and SciPy. QuTiP was used to construct the relevant single-mode quantum states and operators in a truncated Fock basis, including displaced squeezed states, coherent states, thermal states, and the corresponding transformed states after displacement, inverse squeezing, and noisy channels.

For the phase-diffusion model, the mixed received states $\rho_i^\text{pd}$ were generated by averaging over random phase rotations weighted by the Gaussian phase-noise distribution, see Eq.~\eqref{eq:states_phaseDiffuse_psi}. For the thermal-noise model, the states $\zeta_i'$ entering the PNR(M) were obtained from the OOK displacement (i.e. $D(0)$ and $D(2\gamma)$) representation of the thermal state $\rho_\text{th}(n_t)$, see Eq.~\eqref{eq:zeta_thermal_mixture_02}. For the pure-loss channel, the received states $\rho_i^\text{loss}$ were modeled as BPSK displaced squeezed thermal states, see Eq.~\eqref{eq:rho_pureLoss}. In all noisy cases, the conditional counting statistics were evaluated numerically and inserted into the MAP rule to determine the optimal decision and corresponding error probability.

The photon-number cutoff $N_\text{cut}$ in the Fock basis was chosen sufficiently large to ensure numerical convergence for all reported parameter ranges. In this paper, the simulations use a cutoff $N_\text{cut} = 100$. We verified numerical convergence by increasing the cutoff dimension until the resulting error probabilities changed negligibly. The same convergence criterion was applied to the calculated Wigner functions and Fock-basis populations shown in the figures.

\bibliography{refs}

\section*{Acknowledgements}
This work was supported by 
Independent Innovation Science Fund Program of National University of Defense Technology, China (Grant No.~22-ZZCX-036), 
Key research\&development program of Guangxi (Grant No.~GuiKeAB23075155), 
Innovative Talent Development Fund of Information Support Force Engineering University (Grant No.~XJKT-QT-25-02-GW), 
and the Key Research \& Development Program of Guangxi (Grant No.~AB23075112). 

\section*{Author contributions}
E. B. conceived the study, performed the theoretical analysis and numerical simulations, and wrote the initial manuscript draft. J. P. and T. W. provided critical guidance on the research direction and methodology. C. D. and Z. Z. supervised the project, interpreted the results, and secured the funding. T. W., J. P., K. W., F. S., C. Z. and Y. L. offered expertise in specialized theoretical aspects and reviewed the manuscript.

\section*{Data Availability}
The data that support the findings of this study are available from the authors upon request.
\section*{Code Availability}
The numerical simulations in this study were performed in Python using QuTiP 5.1.1. The custom scripts used to generate the results are available from the corresponding author on reasonable request.
\section*{Competing interests}
The Authors declare no Competing Financial or Non-Financial Interests.

\end{document}